%% file: main.tex
\documentclass[twocolumn, 
                floatfix]{aastex631}
\usepackage{amsmath}
\received{August 31 2023}
\revised{October 4 2023}
\accepted{October 16 2023}

\submitjournal{ApJL}

\shorttitle{`Wrong Way' Radio Relics}
\shortauthors{B\"oss et al.}

\begin{document}

\title{A formation mechanism for `Wrong Way' Radio Relics}

\correspondingauthor{Ludwig M. B\"oss}
\email{lboess@usm.lmu.de}

\author[0000-0003-4690-2774]{Ludwig M. B\"oss}
\affiliation{Universitäts-Sternwarte, Fakultät für Physik, Ludwig-Maximilians-Universität München, Scheinerstr.1, 81679 München, Germany }
\affiliation{Excellence Cluster ORIGINS, Boltzmannstr. 2, D-85748, Garching, Germany}

\author[0000-0001-8867-5026]{Ulrich P. Steinwandel}
\affiliation{Center for Computational Astrophysics, Flatiron Institute, 162 5th Avenue, New York, NY 10010}

\author{Klaus Dolag}
\affiliation{Universitäts-Sternwarte, Fakultät für Physik, Ludwig-Maximilians-Universität München, Scheinerstr.1, 81679 München, Germany }
\affiliation{Max Planck Institute for Astrophysics, Karl-Schwarzschildstr. 1, D-85748, Garching, Germany}



\begin{abstract}
Radio Relics are typically found to be arc-like regions of synchrotron emission in the outskirts of merging galaxy clusters, bowing out from the cluster center.
In most cases they show synchrotron spectra that steepen towards the cluster center, indicating that they are caused by relativistic electrons being accelerated at outwards traveling merger shocks.
A number of radio relics break with this ideal picture and show morphologies that are bent the opposite way and show spectral index distributions which do not follow expectations from the ideal picture.
We propose that these `Wrong Way' Relics can form when an outwards travelling shock wave is bent inwards by an in-falling galaxy cluster or group.
We test this in an ultra-high resolution zoom-in simulation of a massive galaxy cluster with an on-the-fly spectral Cosmic Ray model.
This allows us to study not only the synchrotron emission at colliding shocks, but also their synchrotron spectra to adress the open question of relics with strongly varying spectral indices over the relic surface.
\end{abstract}



\input{intro}

\input{methods}

\input{results}

\input{discussion}

\input{conclusion}
%
\section*{Acknowledgments}
We thank the anonymous referee for their helpful and constructive referee report, which improved the quality of this manuscript.
We thank Nadia Böss for the illustration of Fig. 1.
LMB would like to thank Bärbel Koribalski, Sebastian Nuza and Harald Lesch for helpful discussion.
LMB and KD are acknowledging support by the Deutsche Forschungsgemeinschaft (DFG, German Research Foundation) under Germanys Excellence Strategy -- EXC-2094 -- 390783311 and support for the COMPLEX project from the European Research Council (ERC) under the European Union’s Horizon 2020 research and innovation program grant agreement ERC-2019-AdG 860744.
 UPS is supported by the Simons Foundation through a Flatiron Research Fellowship at the Center for Computational Astrophysics of the Flatiron Institute. The Flatiron Institute is supported by the Simons Foundation.

%

\facilities{ The simulations were carried out at the CCA clusters ``rusty'' located in New York City as well as the cluster ``popeye'' located at the San Diego Supercomputing center (SDSC).
 Additional computations were carried out at the \textsc{c2pap} cluster at the Leibnitz Rechenzentrum under the project pn36ze.}


\software{Simulations are carried out with \textsc{OpenGadget3} \citep[][]{Springel2005, Dolag2009, Beck2016, Groth2023}. Processing the simulation output was done with \textsc{GadgetIO.jl} \citep[][]{GadgetIO} and \textsc{GadgetUnits.jl} \citep[][]{GadgetUnits}. Mapping SPH data to a cartesian grid was performed with \textsc{SPHKernels.jl} \citep[][]{SPHKernels} and \textsc{SPHtoGrid.jl} \citep[][]{SPHtoGrid}. These packages use the \textsc{julia} programming language \citep[][]{Bezanson2014}.
Figures were generated with \texttt{matplotlib} \citep[][]{Hunter2007}, using \textsc{PyPlot.jl}.
The analysis scripts to reproduce this work are available at Zenodo via \citet{WWR_repo}.
Data of the relevant simulation domain is available at Zenodo via \citet{WWRs_data}.}


\bibliography{main}{}
\bibliographystyle{aasjournal}

\input{appendix}


\end{document}

%% file: intro.tex
\section{Introduction} \label{sec:intro}

Radio Relics are roughly Mpc-size regions of radio emission in galaxy clusters, typically with an arc-like morphology, which shows strong polarisation \citep[typically $\sim 30\%$ at 1.4 GHz, e.g.][]{Rajpurohit2022}, steep integrated radio spectra ($L_\nu\propto \nu^\alpha$, where $\alpha \sim -1.1$) and a steepening of these spectra towards the cluster center \citep[see][for a recent review]{Weeren2019}.
They are typically associated with ongoing mergers between massive galaxy clusters \citep[see e.g.][]{Ensslin1998, Roettiger1999, Ensslin2002, Brueggen2012, Brunetti2014}.
These mergers dissipate a large fraction of their potential energy in the form of shocks which heat the intra-cluster medium (ICM) to $\sim 10^8$ K.
This can be observed as thermal X-ray emission of the fully ionized plasma \citep[e.g.][for a review]{Boehringer2010}.
A smaller part of the shock energy is dissipated into the acceleration of Cosmic Ray (CR) electrons and protons in a process called ``diffusive shock acceleration" \citep[DSA, see e.g.,][the latter for a review]{Bell1978a, Bell1978b, Blandford1987, Drury1983}{}{}.
In this process (supra-)thermal particles cross a shock front and are scattered by MHD turbulence from the downstream of the shock back into the upstream.
They gain energy at every crossing until their gyro-radii are large enough to escape from the acceleration region or they are advected away in the downstream of the shock.
Hybrid and PIC plasma simulations of shock fronts show that this process can efficiently accelerate protons in low-$\beta$ supernova shocks \citep[e.g.][the latter for a review]{Caprioli2014, Caprioli2018, Caprioli2020, Pohl2020} and high-$\beta$ structure formation shocks \citep[e.g.,][]{Ryu2019, Ha2023}.
For electrons it is found that this process is harder to trigger, as their gyro-radii are smaller at equivalent magnetic field strength and with that it is more difficult to start a cyclical DSA process and with that efficient acceleration of thermal electrons to the GeV energies expected from synchrotron emission by radio relics.
They require an efficient pre-acceleration process such as (stochastic) shock-drift acceleration (SDA), or a seed population stabilized against cooling, to efficiently part-take in a DSA process \citep[see e.g.,][]{Guo2014, Park2015, Kang2019, Tran2020, Kobzar2021, Amano2022, Tran2023}.
On top of that the acceleration efficiency is found to be dependent on the shock obliquity, the angle between shock propagation and magnetic field vector.
Typically it is found that protons are more efficiently accelerated at quasi-parallel shocks \citep[see e.g.,][]{Kang2013, Caprioli2014, Ryu2019}, while electrons are more efficiently accelerated at quasi-perpendicular shocks \citep[e.g.,][]{Guo2014, Kang2019, Ha2021, Amano2022}.
The results from small-scale plasma simulations have been adopted in cosmological simulations to model emission originating from structure formation shocks \citep[see e.g.,][]{Hoeft2008, Pfrommer2008b, Pfrommer2007, Pfrommer2008, Pfrommer2017, Skillman2013, Vazza2012, Vazza2016, Wittor2017, Banfi2020, Wittor2021, Ha2023}.
However, the efficiencies found in plasma simulations are not sufficient to explain the high synchrotron brightness of radio relics \citep[see][for a recent discussion]{Botteon2020}.\\
Recent observations of radio relics show not only bright arc-like structures, but also more complex morphologies such as S-shapes \citep[e.g.,][]{deGasperin2022}, flat relics with varying thickness \citep[e.g.,][]{Weeren2016toothbrush, Rajpurohit2020, Rajpurohit2020a} and filamentary stuctures \citep[e.g.][]{Trasatti2015, Rajpurohit2022a, Chibueze2023}.
There is also a small set of radio relics that show a curvature which points in the ``wrong" direction.
Instead of the typical outward-bent (convex) shape of the relic, away from the cluster center, they show an inward-bent (concave) morphology.
Examples of these relics can be found in the \textit{Ant Cluster} (\textsc{PSZ2 G145.92-12.53}) \citep[][]{Botteon2021}{}{}, PSZ2 G186.99+38.65 \citep[][]{Botteon2022a}, \textit{Source D1} in Abell 3266 \citep[][]{Riseley2022}, \textsc{SPT-CL J2023-5535} \citep[][]{HyeongHan2020}, Abell 168 \citep[][]{Dwarakanath2018} and the southern relic in \textsc{Ciza J2242.8+5301} \citep[e.g.,][]{Weeren2010, Stroe2013, Stroe2016, Hoang2017, DiGennaro2018}.
They can show steep synchrotron spectra which would indicate Mach numbers of the underlying shocks that are in disagreement with the critical Mach numbers found to be required to efficiently accelerate CR electrons \citep[e.g.,][]{Kang2019} and some of their spectra are better fit by broken power-laws rather than a single one hinting towards overlapping (re-)acceleration processes as shown in \citet{Riseley2022} and inititally also in \citet{Owen2014, Trasatti2015, Parekh2022}.
However, follow-up observations indicate that there is no spectral break in the majority of these relics after all \citep[][]{Benson2017, Rajpurohit2022}, making the source \textit{D1} the only relic so far that shows this peculiar spectral behaviour.
The southern relic of \textsc{Ciza J2242.8+5301} shows additional strong variations of the synchrotron slope, which makes it hard to explain in the context of DSA at a single shock front \citep[see discussion in][]{DiGennaro2018}.
Cosmological simulations of galaxy clusters show that mergers between clusters are not isolated events and that merger shocks can deform as they expand into highly complex and turbulent ICM \citep[e.g.][]{Hoeft2008, Skillman2013, Wittor2017, Nuza2017}.
In this work we propose that a possible formation mechanism for these `Wrong Way' relics \citep[as they are referred to in][]{Riseley2022} is the collision of an outwards travelling shock front with an in-falling substructure.
We investigate this scenario in the sibling simulation of an ultra-high resolution MHD simulation of a $M_\mathrm{vir} \approx 1.3 \times 10^{15} M_\odot$ galaxy cluster introduced in \citet{250xMHD}, where we attached a population of CR protons and electrons to every resolution element of our simulation.
This effectively turns every particle into a tracer particle for CRs, while also accounting for feedback by the CR component on the thermal gas. 
We resolve these populations with a spectral resolution of 12 bins for protons, and 48 bins for electrons over a range of 6 orders of magnitude in momentum.
The distribution function of the CRs is updated on-the-fly at every timestep of the simulation according to the method presented in \citet{Boess2023}.
This allows us to study CR electron injection at colliding shocks and the subsequent cooling of the relativistic electron population.
To the best of our knowledge this simulation is the first of its kind.
This work is structured as follows:
In Sec.~\ref{sec:methods} we describe the simulation code, CR model and initial conditions used in this work.
In Sec.~\ref{sec:results} we study the `Wrong Way' Relic (WWR) found in the simulation and its origin.
Sec.~\ref{sec:discussion} contains a discussion of our findings and a comparison to observed systems.
Finally Sec.~\ref{sec:conclusion} contains our conclusion and outlook to future work.

%% file: methods.tex
\section{Methods \label{sec:methods}} 

The simulation used in this work was carried out with the Tree-SPMHD code \textsc{OpenGadget3}.
\textsc{OpenGadget3} is a derivative of \textsc{Gadget2} \citep{Springel2005} with improvements to the hydro and gravity solvers as well as additional physics modules.
The SPH solver is updated as described in \citet{Beck2016} to include higher order kernels and their bias correction \citep[see][]{Dehnen2012} and artificial viscosity as well as physical conduction to improve the mixing behavior and shock capturing of SPH \citep[e.g.][]{Price2012, Hopkins2013, Hu2014}.
Magnetohydrodynamics (MHD) have been implementation by \citet{Dolag2009} with updates to include non-ideal MHD in the form of constant (physical) diffusion and dissipation presented in \citet{Bonafede2011}.
Conduction is modelled via a conjugate gradient solver \citep{Petkova2009,Arth2014,SteinwandelSNpaper}, with a suppression factor of the Spitzer value for conduction of 5 per cent.
We adopt a Wendland C4 kernel \citep[][]{Wendland1995, Wendland2005} with 200 neighbors and bias correction as suggested by \citet{Dehnen2012}.
\\
We employ the on-the-fly spectral CR model \textsc{Crescendo} introduced in \citet{Boess2023} to model the time evolution of CR protons and electrons in every resolution element of our simulation.
The time evolution of distributions of CRs in the absence of CR transport, diffusion in momentum space and catastrophic losses can be described by 
\begin{align}
	\frac{D f(p,\mathbf{x},t)}{Dt} &=  \left( \frac{1}{3} \nabla \cdot \mathbf{u} \right) p \frac{\partial f(p,\mathbf{x},t)}{\partial p} \label{eq:fp-adiabatic} \\
	&+ \frac{1}{p^2} \frac{\partial}{\partial p } \left( p^2 \sum_l b_l f(p,\mathbf{x},t) \right) \label{eq:fp-rad}\\
	&+ j(\mathbf{x}, p, t), \label{eq:fp-sources}
\end{align}
where we used $\frac{Df}{Dt} = \frac{\partial f}{\partial t} + \mathbf{u} \cdot \nabla f$ due to \textsc{OpenGadget3} being a Lagrangian code.
The right side of Eq.~\ref{eq:fp-adiabatic} describes changes due to adiabatic compression or expansion of the gas the CRs are confined in, Eq. \ref{eq:fp-rad} describes energy losses and Eq. \ref{eq:fp-sources} is the source term.
We represent $f(p,\mathbf{x},t)$ as piece-wise powerlaws in momentum space with 2 bins/dex for protons and 8 bins/dex for electrons in the dimensionless momentum range $\hat{p} \equiv \frac{p_i}{m_i c} \in [0.1, 10^5]$, where $p_i$ and $m_i$ refer to the momentum and mass for protons and electrons, respectively.
The distribution function is updated at every timestep following the two-moment approach as introduced in \citet{Miniati2001} by computing CR number and energy changes per bin.
Adiabatic changes are accounted for at every timestep via the density change within a SPH particle.
We model energy losses of electrons due to synchrotron emission and inverse Compton scattering off CMB photons.
As a source term we employ the DSA parametrisation by \citet{Kang2013} for the dependency on sonic Mach number ($\eta(\mathcal{M}_s)$), which allows for DSA at shocks beyond a critical Mach number $\mathcal{M}_s > 2$ and saturates at a maximum efficiency of $\eta_\mathrm{max} \approx 0.2$.
In addition to that we use the model by \citet{Pais2018} for the dependency of CR acceleration efficiency on shock obliquity ($\eta(\theta_B)$).
Ultimately we divert a fraction 
\begin{equation}
    \eta_\mathrm{tot} = \eta(\mathcal{M}_s) \times  \eta(\theta_B)
\end{equation}
of the entropy change over the shock into the CR component.
We detect the shock properties on-the-fly in the simulation with the shock finder introduced by \citet{Beck2016_shock} with improvements to compute the shock obliquity as the angle between the pressure gradient within the kernel (which we treat as the shock normal $\hat{\mathbf{n}}$) and the magnetic field vector upstream of the shock $\mathbf{B}_u$.
The slope of the injected CR spectrum follows linear DSA theory and we use a fixed electron to proton injection ratio of $K_{e/p} = 0.01$.
The CR component exerts feedback on the thermal gas by solving the pressure integral
\begin{equation}
        P_{\mathrm{CR},c} = \frac{4\pi \: c}{3} \: a^{4} \: \int\limits_{p_{\mathrm{min}}}^{p_{\mathrm{cut}}} dp \: p^3 f(p) 
    \label{eq:pressure_integral}
\end{equation}
between the minimum momentum $p_\mathrm{min}$ represented by the CR population and the cutoff of the distribution function $p_\mathrm{cut}$.
We start the CR injection at $z=4$ to avoid too strong time-constraints due to very efficient high-momentum energy losses of CR electrons.\\
Synchrotron emission is calculated directly from the evolved electron distribution function (see Appendix \ref{sec:synch} for details).
\\
We use a zoomed-in initial condition of a massive galaxy cluster with a virial mass of $M_\mathrm{vir} \approx 1.3 \times 10^{15}~M_\odot$ from the sample presented in \citet{Bonafede2011}.
The cluster is up-sampled to 250x base resolution, which corresponds to a mass resolution of $M_\mathrm{gas} \approx 8.7 \times 10^5 M_\odot$ and $M_\mathrm{DM} \approx 4.7 \times 10^6 M_\odot$ for gas and dark matter particles, respectively.
We reach a maximum resolution for a gas particle of $h_\mathrm{sml, min} \approx 1$ kpc with a gravitational softening of $\epsilon = 0.48 \: h^{-1} c$ kpc.
The cluster was selected from a lower-resolution dark matter-only simulation of a Gpc volume, which is large enough to provide a large sample of systems above a few $10^{15} M_\odot$.
The parent simulation used a WMAP7 cosmology with $\Omega_{0} = 0.24$, $\Omega_{\Lambda} = 0.76$, $\Omega_\mathrm{baryon} = 0.04$, $h = 0.72$ and $\sigma_{8} = 0.8$, which we also adopt for the present simulation.
We start the simulation at redshift $z=310$ and seed a constant magnetic field in x-direction with $B_0 = $10$^{-14}$ G \citep[see][for a study of the impact of the choice of $B_0$]{SteinwandelClusterpaper}.
The initial conditions of this cluster at this resolution have been used to study the interaction between internal- and accretion shocks in \citet{Zhang2020a, Zhang2020b} and its magnetic field has been studied in \citet{250xMHD}.
\\

%% file: results.tex
\section{Results \label{sec:results}}

\subsection{Merger Geometry}
\begin{figure}
    \centering
    \includegraphics[width=0.42\textwidth]{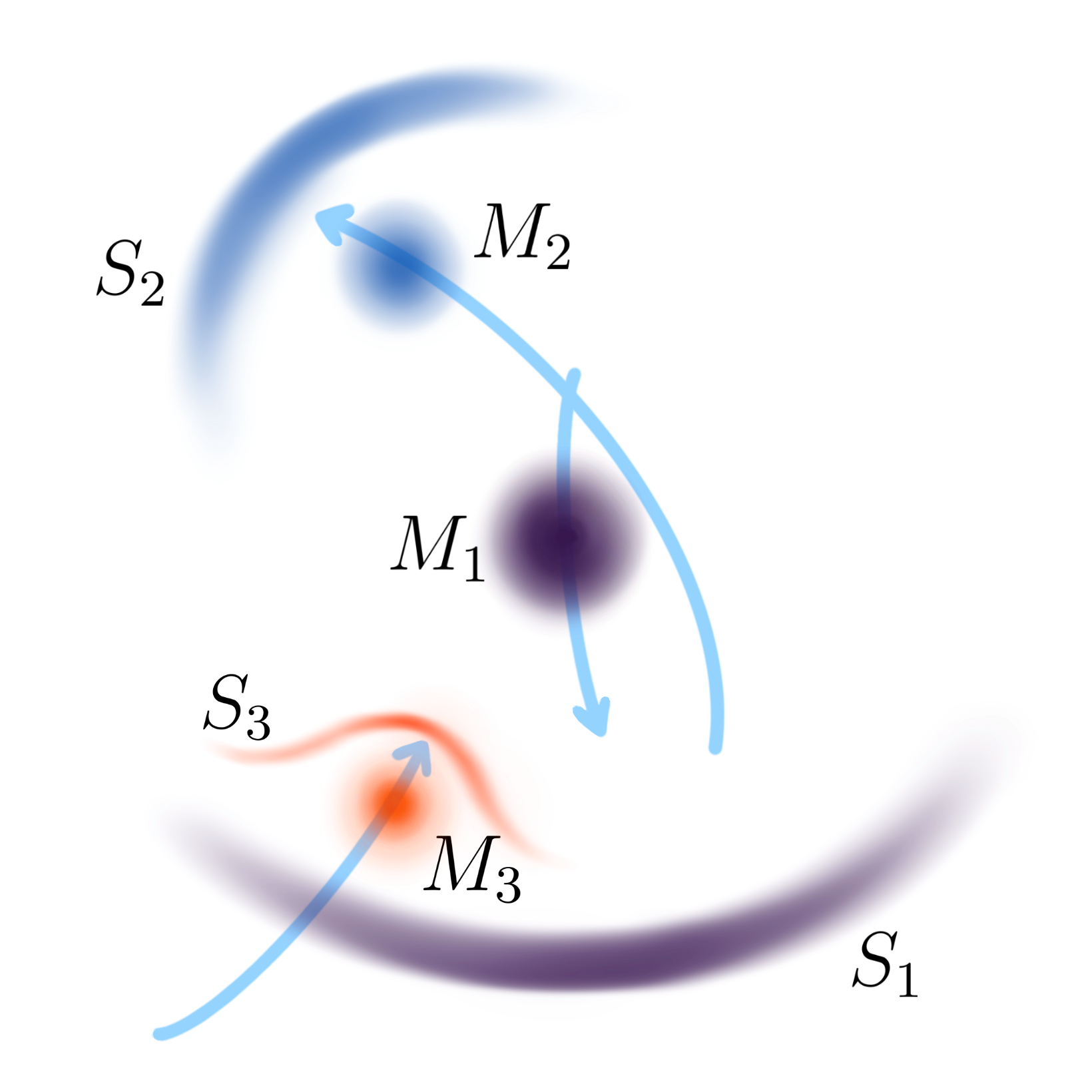}
    \caption{A simplified schematic of the merger geometry that produces the `Wrong Way' relic. The initial merger between $M_1$ and $M_2$ drives two shocks, the weaker of which is subsequently impacted by a third substructure $M_3$. This impact deforms parts the outwards traveling shock $S_1$ and produces the WWR $S_3$.}
    \label{fig:merger_schematic}
\end{figure}
The `Wrong Way' relic in our simulation originates from a triple-merger at $z \sim 0.35 - 0.2$.
We show the schematic of the merger geometry in Fig. \ref{fig:merger_schematic}.
A high-velocity merger with a 1:10 mass ratio between impactor ($M_2 \approx 10^{14} M_\odot$) and target ($M_1\approx 10^{15} M_\odot$) with a large impact parameter of $b \approx 500$ kpc drives two shock waves.
These shocks follow the canonical picture \citep[e.g. Fig.~7 in][]{Weeren2019} of the lighter merging partner ($M_2$) driving a strong bow-shock ($S_2$ in our schematic), while the heavier merging partner ($M_1$) drives a weaker counter shock ($S_1$) in the in-fall direction of the lighter partner.
This counter shock is subsequently impacted by a third merger partner ($M_3$), a group of galaxies with a total mass of $M_3 \approx 2\times 10^{13} M_\odot$, which ultimately passes through the shock surface and falls into the larger merger partner ($M_1$) in a low-impact parameter merger with $b \approx 35$ kpc.
The impact of the group deforms the weaker counter shock ($S_1$) first from a convex shape at $z=0.32$ to a concave shape at $z=0.29$ and subsequently to a v-like shape pointing towards the cluster center at $z=0.27$, which also leads to a complex superposition of the different parts original shock surface with different mach numbers as well as differently aged cosmic ray electron population.
\\
Due to our system being a single, isolated cluster we cannot make any predictions for the minimum critical mass of an in-falling sub-structure that is able to deform such a shock front, or the statistical frequency of such an event.
We leave this question for future work with cosmological boxes, to allow for a statistical analysis.
\subsection{The Simulated `Wrong Way' Radio Relic}
\begin{figure*}
    \centering
    \includegraphics[width=\textwidth]{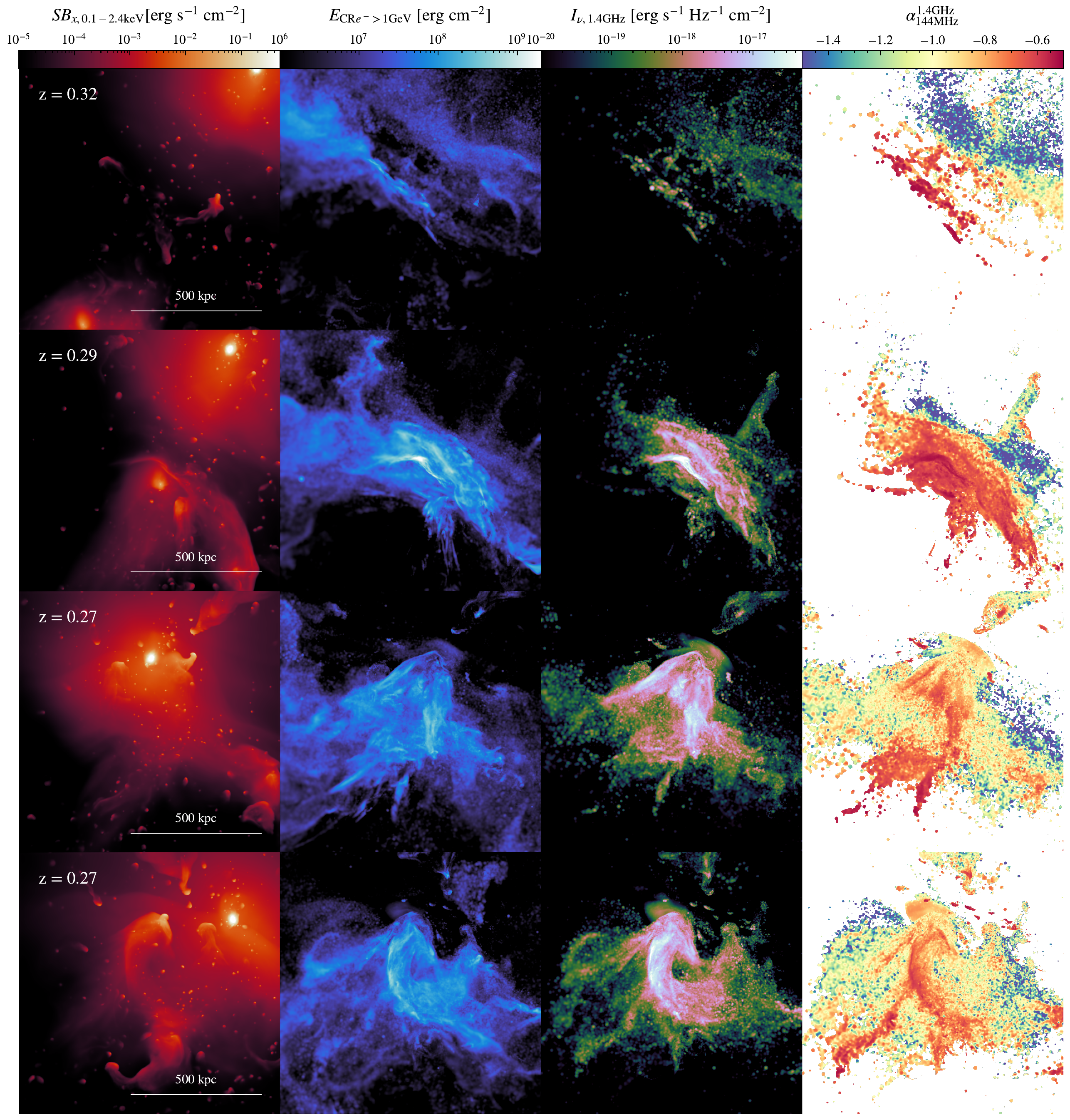}
    \caption{From left to right: X-ray surface brightness, CR electron energy of electrons with $E > 1$ GeV, synchrotron surface brightness at 1.4 GHz and the slope of the synchrotron spectrum between 144 MHz and 1.4 GHz.
    The upper three rows show the time evolution of the in-falling group in the $xz$-plane of the simulation, the lowest row shows the same relic at $z = 0.27$ in the $yz$-plane. To obtain the images the SPH data is mapped to a grid with a resolution of $\Delta_\mathrm{pix} \approx 1$ kpc, which corresponds to a resolution of $\theta_\mathrm{pix} \approx 0.24"$ at $z=0.27$}.
    \label{fig:relic}
\end{figure*}
\begin{figure}
    \centering
    \includegraphics[width=0.48\textwidth]{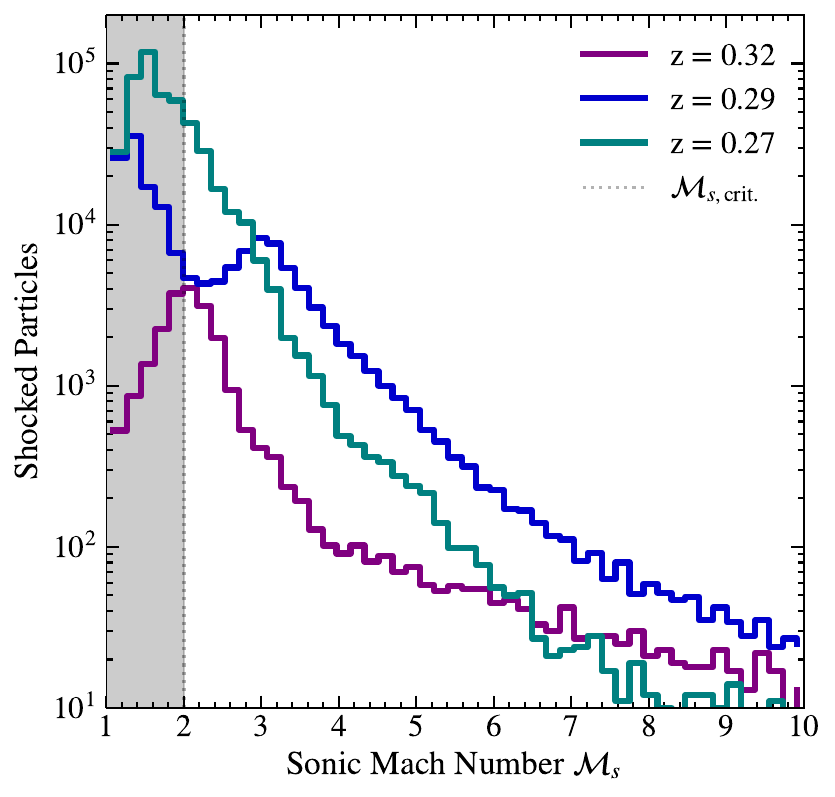}
    \caption{Histograms of the sonic Mach number $\mathcal{M}_s$ for the three output times shown in Fig.~\ref{fig:relic}. The colors correspond to the different times and the dotted line indicate the critical Mach number beyond which CR (re-)acceleration can occur in our model.}
    \label{fig:mach}
\end{figure}
\begin{figure}
    \centering
    \includegraphics[width=0.49\textwidth]{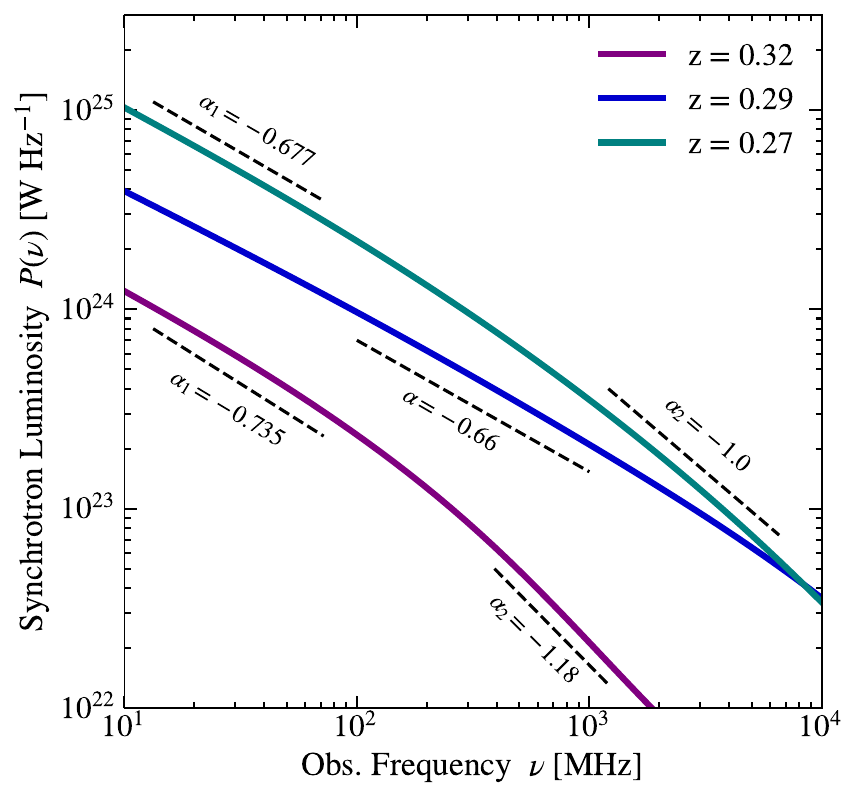}
    \caption{Time evolution of the synchrotron spectrum. Colors correspond to the times in Fig. \ref{fig:relic} and \ref{fig:mach}. Dashed lines and labels show the spectral slope in the indicated frequency ranges.}
    \label{fig:spectral_evolution}
\end{figure}
\begin{figure*}
    \centering
    \includegraphics[width=\textwidth]{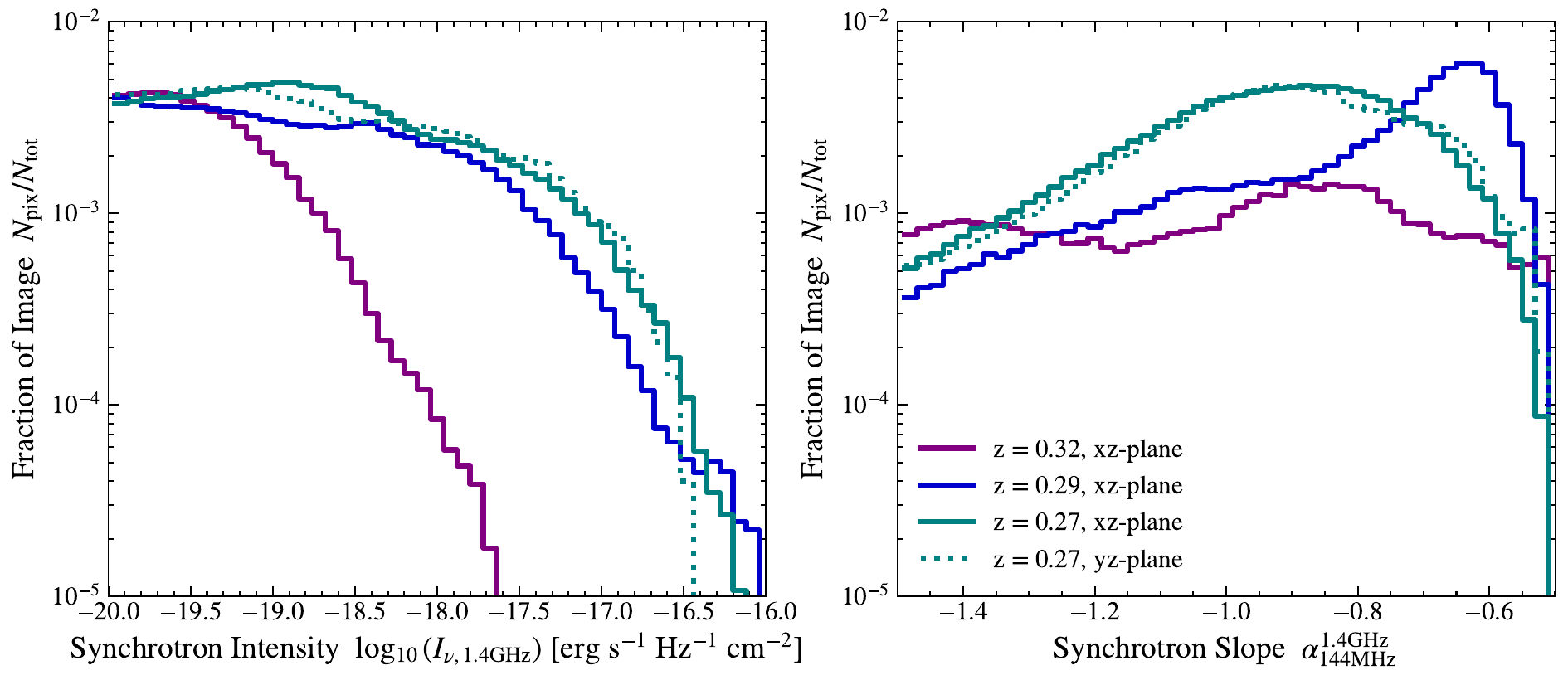}
    \caption{Histograms of the synchrotron surface brightness \textit{(left)} and spectral slope \textit{(right)} obtained from the images in Fig. \ref{fig:relic}. As before colors correspond to the times in Fig. \ref{fig:relic} and \ref{fig:mach}. The dotted line indicates the same relic at $z=0.27$ in the $yz$-plane of the simulation, as in the lowest row of Fig. \ref{fig:relic}.}
    \label{fig:map_histograms}
\end{figure*}
Fig.~\ref{fig:relic} from top to bottom shows the time evolution of the counter shock $S_1$ in the $xz$-plane of the simulation and its phasing through morphologies matching various `Wrong Way' relics.
The bottom row shows the same relic as the row above in the $yz$-plane.
From left to right we show the X-ray surface brightness, CR electron energy contained in the part of the potentially synchrotron bright population with $E > 1$ GeV, the synchrotron surface brightness at 1.4 GHz and the slope of the synchrotron spectrum obtained by fitting a powerlaw to the surface brightness at 144 MHz and 1.4 GHz.
These images are obtained by mapping the SPH data to a 2D grid following the algorithm described in \citet{Dolag2005} with a pixel-size of $\Delta_\mathrm{pix} \approx 1$ kpc.
This corresponds to a resolution of $\theta_\mathrm{pix} \approx 0.24"$ at $z=0.27$ and with that is significantly below current observational limits.
Accompanying to that we show the distribution of sonic Mach number $\mathcal{M}_s$ of the different panels of Fig. \ref{fig:relic} in Fig. \ref{fig:mach} and the synchrotron spectra in Fig. \ref{fig:spectral_evolution}.
In Fig.~\ref{fig:map_histograms} we show the historgrams of pixels in Fig.~\ref{fig:relic} as a function of synchrotron intensity and spectral slope.
\\
At $z=0.32$, in the top row of Fig. \ref{fig:relic}, we see the acceleration of CR electrons at the counter shock of the main merger event.
Fig. \ref{fig:mach} shows that only a fraction of the shocked particles are above the critical Mach number $\mathcal{M}_{s,\mathrm{crit}} = 2$ and with that can accelerate CRs.
We can readily identify the part of the shock surface that accelerates CRs in the center of the images, as it is the most synchrotron bright part and shows a relatively flat synchrotron spectrum.
These CRs are accelerated at the contact surface between outwards traveling shock and the atmosphere of the in-falling halo.
The steeper parts of the spectrum in the upper right corner of the images indicate that these electrons have been accelerated at earlier times of the shock propagation and have been freely cooling since.
This is also evident in the synchrotron spectrum in Fig. \ref{fig:spectral_evolution}, which shows a strong break above $\nu \sim 200$ MHz.
The counter shock is initially not very synchrotron bright, akin to the counter-shock in Abell 3667 \citep[e.g.][]{deGasperin2022} or \textsc{CIZA J2242.8+5301} \citep[][]{DiGennaro2018}.
\\
At $z=0.29$ the collision between outwards traveling counter shock and the bow-shock of the in-falling group ($M_3$) increases the sonic Mach number and with that the acceleration efficiency of the shock \citep[see e.g.][for studies of multi-shock scenarios]{Kang2021, Inchingolo2022}.
Fig. \ref{fig:mach} shows that while the majority of the shocked particles remain sub-critical, the shock develops a second Mach number peak around Mach 3.
This significantly increases the synchrotron surface brightness at the contact surface of the shocks, flattens the synchrotron spectrum to almost the the theoretical limit of DSA and erases the spectral break.
A spectral slope of $\alpha_{100 \: \mathrm{MHz}}^{1 \: \mathrm{GHz}} = -0.66$ indicates $\mathcal{M}_s \approx 3.6$, in good agreement with the underlying Mach number distribution.
This injection domination can also be seen in Fig.~\ref{fig:map_histograms} where the images at $z=0.29$ show a strong bump in synchrotron slopes between $\vert\alpha\vert \approx 0.7 - 0.55$ and a small bump in synchrotron intensity around $I_\nu \approx 10^{-16.5} - 10^{-16}$ erg s$^{-1}$ Hz$^{-1}$ cm$^{-2}$.\\
The in-falling sub-structure deforms the outwards traveling shock towards a relic pointing ``the wrong way", similar to the source \textit{D1} observed by \citet{Riseley2022}.
In the case of our relic the flat spectrum part is further extended, which we attribute to the shock being further bent inwards, compared to \textit{D1}.\\
In Fig.~\ref{fig:relic_rot} we rotate the image into the merger plane and can see how the aged, steep-spectrum population disappears behind the newly injected electrons at the inward bent relic.
Comparing to the same rotation at $z=0.32$ indicates that the best morphological fit to \textit{D1} would lie between $z=0.32$ and $z=0.29$, however there is no output available at this time.\\
The collision between shock waves is also visible in our X-ray image (left panel, second row in Fig. \ref{fig:relic}), which matches the detection of a shock in X-ray by \citet{Sanders2022}.
The in-fall scenario proposed here also produces a radio relic-like structure within $r_{500}$, which is unlikely in the classical picture of radio relics \citep[e.g.,][]{Vazza2012}.\\
At $z=0.27$, as the in-falling halo passes through the outwards traveling shock its own bow-shock collides with the older shock, causing the relic to deform further into a v-shaped morphology, such as in the counter shock to the \textit{sausage relic} \citep[e.g.][]{Stroe2013, DiGennaro2018}, or the relic in Abell 2256 \citep[][]{Rajpurohit2022a}.
The Mach number distribution over the shock surface has become smoother at this point, with the bulk of the shock being sub-critical, however the total number of particles with $\mathcal{M}_s > 2$ has increased compared to the relic at $z=0.29$.
This leads to efficient acceleration at a part of the shock surface, visible in increased synchrotron surface brightness and flatter synchrotron spectra.\\
In general the relic is however cooling and adiabatic compression dominated.
This becomes visible in Fig.~\ref{fig:map_histograms} where synchrotron intensity is increased in the $I_\nu \sim 10^{-17.5} - 10^{-16.5}$ erg s$^{-1}$ Hz$^{-1}$ cm$^{-2}$ range.
However, spectra are generally steeper, indicating that the increase in intensity is partly by injection and partly by adiabatic compression of an already cooling electron population.\\
A morphological best match for the relic in Abell 2256 is expected to lie between $z=0.29-0.27$ shown here, however the simulation output for this time is not available.
For the lower panels of Fig.~\ref{fig:relic} we rotate the image by $90^\circ$, as this projection more closely resembles the observations of \citet{DiGennaro2018}.
The collision of two shocks as shown here leads to a superposition of multiple DSA-like events due to strong variations of the Mach number over the shock surface.
This leads to strong variations of synchrotron surface brightness and spectral shape between the regions of the shock surface where efficient (re-) acceleration can take place and the regions that are dominated by cooling and adiabatic compression.\\
These variations can also be seen in the integrated spectrum in Fig. \ref{fig:spectral_evolution}, where the lower frequency end of the spectrum is strongly injection dominated and the high frequency end of the spectrum shows a significant steepening beyond $\nu \sim 1$ GHz in the cooling dominated part.
This result is valid for the two lower panels of Fig.~\ref{fig:relic}, as we are dealing with integrated quantities. We have confirmed this by comparing the integrated spectrum obtained based on the data directly from the SPH particles as well as integrated maps under three different projections.
We find no qualitative difference between these approaches.

%% file: discussion.tex
\section{Discussion \label{sec:discussion}}
To discuss our findings we will compare the morphologies in chronological order to similar observed systems.
Albeit the number of observed WWRs is still small, the recent discoveries due to \textsc{ASKAP} and MeerKAT indicate that with increased sensitivity a number of new WWRs can be detected over time.

\subsection{Abell 520}

Before the onset of the WWR morphology our cluster undergoes an internal merger with an in-falling group in the cluster periphery.
This group falls into the cluster at a similar trajectory as the cluster driving the current shock waves and is therefore in the path of the weaker counter-shock of the ongoing merger.
A similar setup is observed in Abell 520 by \citet{Hoang2019}.
They detect a shock with Mach number $\mathcal{M}_\mathrm{SW} = 2.6_{-0.2}^{+0.3}$ propagating in SW direction with a weaker counter-shock moving with $\mathcal{M}_\mathrm{SW} = 2.1 \pm 0.2$ in NE direction.
Along the NE diagonal Chandra observations by \citet{Andrade2017} indicate in-falling matter along a similar path as the ongoing merger.
This shows that the geometric setup is possible, albeit rare and Abell 520 could host a WWR in the future.

\subsection{Abell 3266}
At $z=0.32-0.29$ our WWR resembles the one observed by \cite{Riseley2022}, at a distance of $\sim$ 1 Mpc from the center of Abell 3266.
Their relic is very faint and shows a very steep spectral index of $\alpha = -2.76$ in the part that is observable in frequencies above 1043 MHz.
The lower frequency end of the relic spectrum is significantly flatter, with a spectral index of $\alpha \approx -0.72$.
This indicates that there is a re-acceleration process which is superimposed on an older cooling spectrum, even though the very steep spectrum still poses problems under this assumption \citep[see discussion in][]{Riseley2022}.
Xray observations with eROSITA \citep[][]{Sanders2022} show a number of discrete sources in close proximity to \textit{D1}, but no extended sources that could indicate an infalling group.
The extended sources $X4$ and $X6$ that lie in (projected) close proximity to \textit{D1} have significantly higher photometric redshifts ($z=0.532$ for $X4$ and $z=0.907$ for $X6$) than Abell 3266 \citep[$z=0.0589$,][]{Struble1999}, which shows that these are background sources and not in-falling groups.
However, as can be seen in the left panel of Fig. \ref{fig:relic} at $z=0.27$, depending on the projection it is not necessarily easy to distinguish the in-falling structure in the X-ray emission.

\subsection{PSZ2G145.92-12.53}
Another concave radio relic detected in \textsc{PSZ2G145.92-12.53} similarly shows an increase in X-ray flux with concave morphology in close proximity to the relic \citep[see Fig.~1 in][]{Botteon2021}.
We note that there is also a detected peak in X-ray surface brightness akin to the one observed in the \textit{Rim} region in \textsc{PSZ2G145.92-12.53}, indicating that similar effects may be at play there, as briefly discussed by the authors.

\subsection{Abell 2256}
As previously discussed, at $z=0.27$ in the $xz$ plane, corresponding to the third row in Fig.~\ref{fig:relic}, our WWR closely resembles the steep radio relic found in Abell 2256.
\citet{Rajpurohit2022a} note an association between the relic and the source F, without an X-ray counterpart \citep[see also][]{Owen2014, Ge2020}.
This could hint towards the group having passed the shock before in a similar process as discussed here.
The superposition of injected and cooling parts of the shock surface can also be seen in the color-color plots in \citet{Rajpurohit2022a}, which indicate that the relic consists of a number of overlapping substructures.
The (re-) acceleration of particles in the turbulent downstream of $S_1$, which becomes the upstream of $S_3$, also produces filamentary structures seen in the relativistic electron component (second panels from left in Fig. \ref{fig:relic}) as observed in Abell 2256 \citep[see e.g.][for detailed studies of surface structures in radio relics]{Dominguez-Fernandez2021, Wittor2023}.
The observed relic shows very little spectral steepening, making it difficult to discern if it was bent against its propagation direction.
The little steepening that is being detected however points towards the cluster center akin to our simulated relic, which can indicate a similar process to the one we discussed here.

\subsection{CIZA J2242.8+5301}
In the case of the counter shock to the sausage relic in \textsc{CIZA J2242.8+5301} the reason for the strong variations of synchrotron spectral index is still under debate \citep[see discussion in][]{DiGennaro2018}.
In the context of our merger scenario these variations can be understood as follows: As the outwards traveling shock ($S_1$) collides with the bow-shock of the in-falling substructure ($M_3$) and is deformed, the resulting shock surface ($S_3$) shows strong variations in sonic Mach number.
Wherever the sonic Mach number is $\mathcal{M}_s < 2$ our DSA model allows no CR (re-) acceleration and the pre-existing population is simultaneously cooling due to synchrotron and IC losses and being adiabatically compressed by the sub-critical shock.
This leads to a continuously steepening synchrotron spectrum, while the adiabatic compression leads to an increase in synchrotron surface brightness. 
In regions of the shock surface where $\mathcal{M}_s > 2$ there is ongoing (re-) acceleration of CR electrons, which lead to a flatter spectrum than for the cooled population.
This superposition of cooling- and acceleration dominated areas on the shock surface leads to a strong variation of synchrotron spectral index, as can be seen in the bottom row of Fig.~\ref{fig:relic}.

%% file: conclusion.tex
\section{Conclusion \label{sec:conclusion}}
In this work we showed the first results of a high-resolution simulation of a massive galaxy cluster with an on-the-fly spectral Fokker-Planck solver to study the acceleration, advection and aging of CR electrons in cosmological zoom-in simulations.
We applied this simulation to study a rare form of radio relics that show inward-bent, instead of the typical outward-bent morphologies.
Our results can be summarized as follows:
\begin{itemize}
    \item In complex merging systems with multiple ongoing mergers collisions between bow-shocks of in-falling substructures and outwards traveling merger shocks can deform the outwards traveling shocks in a way that is morphologically very similar to the currently reported \textit{`Wrong Way' relics}.
    \item These collisions between shocks increase the Mach number at the contact surface of the shocks and with that boost the (re-)acceleration efficiency of CR electrons. This makes their detection easier than that of the cooled, outwards moving shock.
    \item The inclusion of an on-the-fly spectral treatment of CR electrons allows to reproduce the large variance of synchrotron spectral slope across the relic surface. This variance stems from the co-existence of an aged CR electron population in the outwards traveling shock and newly injected CRs at the high Mach number regions of the shock surface.
\end{itemize}
Future work will expand our sample size of radio relics by performing further zoom-in simulations of the cluster set presented in \citet{Bonafede2011} at 250x base resolution and will study surface structure and polarisation properties of these relics, as well as $\gamma$-ray emission by the accelerated protons.

%% file: appendix.tex
\appendix

\section{Synchrotron Emission \label{sec:synch}}
\begin{figure*}
    \centering
    \includegraphics[width=\textwidth]{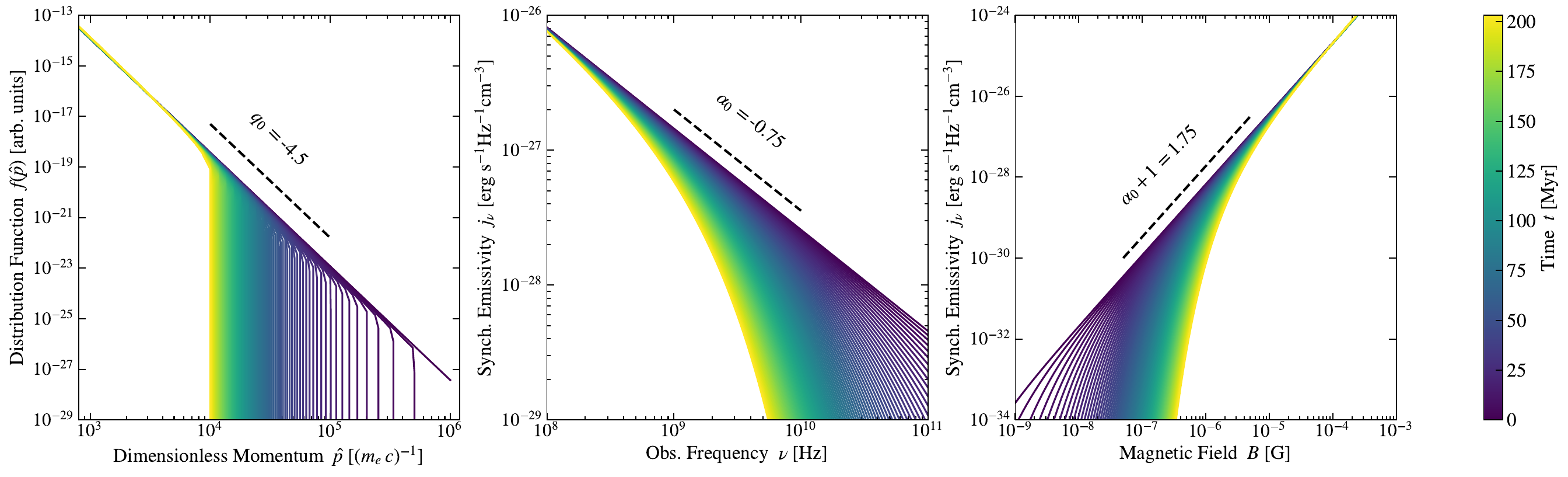}
    \caption{From left to right we show an electron spectrum freely cooling due to IC scattering off CMB photons at $z=0$ and its synchrotron emission scaling with observational frequency (middle) and magnetic field strength (right).}
    \label{fig:synch_scaling}
\end{figure*}
As introduced in \citet{Boess2023}, we compute the synchrotron emission by integrating over the CR electron distribution function and folding it with the synchrotron kernel
\begin{align}
    j_\nu(t) &= \frac{\sqrt{3} e^3}{m_e^2 c^3} \: B(t) \: \sum\limits_{i=1}^{N_\mathrm{bins}} \:\int\limits_0^{\pi/2} d\theta  \sin^2\theta \:  \int\limits_{\hat{p}_\mathrm{i}}^{\hat{p}_\mathrm{i+1}} d\hat{p} \:\: 4\pi \hat{p}^2 f(\hat{p}, t) \: K(x)
    \label{eq:synch_emissivity}
\end{align}
where  $e$ is the elementary charge of an electron, $m_e$ its mass, $c$ the speed of light and $\hat{p}$ the dimensionless momentum.
$K(x)$ is the first synchrotron function
\begin{equation}
    K(x) = x \int_x^{\infty} dz \ K_{5/3}(z)
    \label{eq:synch_kernel}
\end{equation}
using the Bessel function $K_{5/3}$ at a ratio between observation frequency $\nu$ and critical frequency $\nu_c$
\begin{equation}
    x \equiv \frac{\nu}{\nu_c} = \frac{\nu}{C_\mathrm{crit}  B(t) \sin\theta \: \hat{p}^2}; \quad C_\mathrm{crit} = \frac{3e}{4\pi m_e c} \quad .
\end{equation}
The slope of the synchrotron spectrum ($\alpha_0$) scales with the slope $q_0$ of the injected electron spectrum in momentum space as
\begin{equation}
    \alpha_0 = \frac{q-3}{2}
\end{equation}
and with the magnetic field strength as
\begin{equation}
    \alpha_B = \alpha_0 + 1
\end{equation}
In Fig.~\ref{fig:synch_scaling} we show how the synchrotron spectrum of a freely cooling electron spectrum scales with observational frequency and magnetic field strength.
The spectrum (left panel) is set up as if it was injected at a Mach 3 shock and subsequently cools due to IC scattering off CMB photons at $z=0$.
synchrotron losses are ignored, however they would only change the cooling time, not the shape of the spectrum.
The middle panel shows the synchrotron spectrum resulting from the cooling electron spectrum, assuming emission in a magnetic field with strength $B = 5 \:\mu$G.
The right panel shows the emission at 1.4 GHz from the same electron spectrum as a function of magnetic field strength.
We find that we recover the analytic slopes with a relative error below $\Delta\alpha_0 < 0.1\%$ for the injected powerlaw spectrum.
These slopes arise directly from Eq. \ref{eq:synch_emissivity}, without further assumptions or imposed limits.
\section{Relic rotations}
\begin{figure*}
    \centering
    \includegraphics[width=0.8\textwidth]{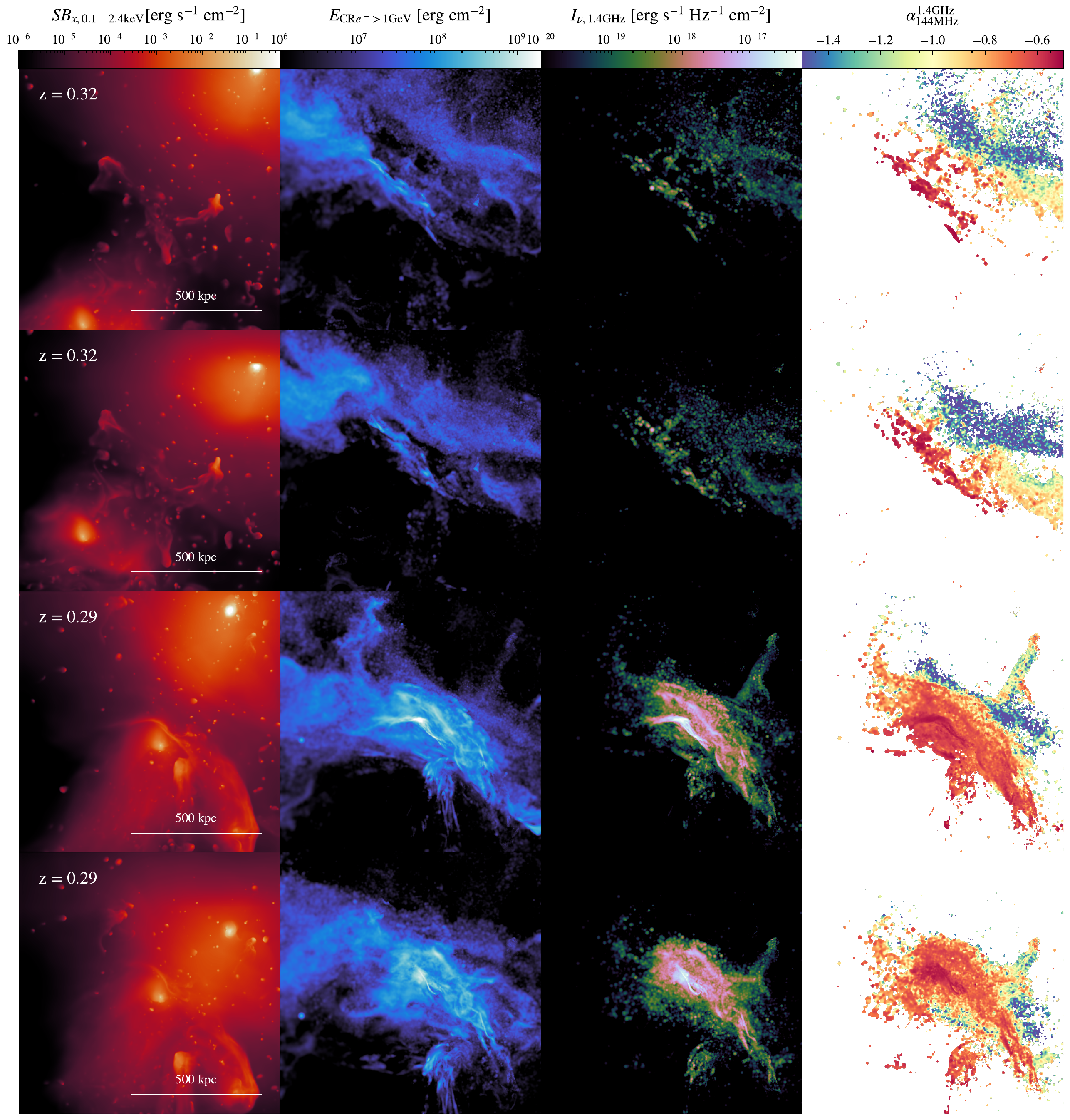}
    \caption{Similar to Fig. \ref{fig:relic} we show our WWR at $z=0.29$ and $z=0.27$
    The first and third row show the default projection in the $xz$-plane, the second and fourth row show the relic rotated into the merger plane. Please note that the lower limit in the left-most column is one order of magntiude lower compared to Fig.~\ref{fig:relic} to improve the visibility of the contact between outwards moving shock and the atmosphere of the in-falling halo.}
    \label{fig:relic_rot}
\end{figure*}
In Fig.~\ref{fig:relic_rot} we rotate the WWR in our simulation into the merger plane.
The panels show a rotation by $30^\circ$ around the $x$-axis from the $xz$-plane.
At $z=0.32$ we can see that the recently injected CRs stem from the first contact between outwards traveling shock and the atmosphere of the in-falling halo.
This leads to a similar morphology as in \citet{Riseley2022}, however the shock is a lot less pronounced due to the early stage of the contact.
At $z=0.29$, as the relic rotates into the merger plane we can see that it is bent further into the cluster center than the \citet{Riseley2022} relic, indicating that it is further progressed than the observed counterpart.
This hides the steep, aged spectrum behind the newly injected electrons, leading to the flat spectrum part dominating the image.
We therefore conclude that an ideal reproduction of source \textit{D1} would lie between $z=0.32-0.29$, most likely shortly after $z=0.32$.